\documentstyle[12pt,oldlfont]{article}
\normalbaselines
\begin{document}
\bibliographystyle{unsrt}

\begin{center}
{\Large {\bf Husimi Parametric Oscillator in Frame of Symplectic Group
and Q-oscillators}}
\end{center}

\begin{center}
V. I. Man'ko\\\ {\em Lebedev Physics Institute, Leninsky pr., 53\\
Moscow 117924, Russia\\}

\end{center}

\begin{abstract}Time--dependent integrals of motion which are linear
forms in position and momentum are discussed for Husimi parametric
forced oscillator. Generalization of these integrals of motion for
q--oscillator is presented. Squeezing and quadrature correlation
phenomena are discussed on the base of Schr\"odinger uncertainty
relation. The properties of the generalized correlated states, squeezed
states, even and odd coherent states (the Schr\"odinger cat states)
are reviewed. The relation of the constructed nonclassical states
to representations of the symplectic symmetry group and finite
symmetry groups is discussed.
\end{abstract}

\section{Q-invariant of Parametric Oscillator}
Husimi \cite{h1} has considered forced parametric oscillator
described by the Schr\"odinger equation
\begin{equation}
i\frac {\partial \Psi (x,t)}{\partial t}
=-\frac {\partial ^{2}\Psi (x,t)}{2\partial x^{2}}
+\frac {\omega ^{2}(t)x^{2}\Psi (x,t)}{2}-f(t)x\Psi (x,t),
\end{equation}
where we take $~\hbar=~m=~\omega (0)=~1$. In the references the Gaussian
packets and analogs of oscillator number states have been found as well
as Green function and transition amplitudes between the energy levels
in a form of series. In \cite{tr69}, \cite{mal70} the integrals of motion
which are linear forms in position and momentum operators have been found
and used to rederive the Husimi solutions. The review of the approach has
been given in \cite{mal79}, \cite{dod89}.

In this talk for the Husimi parametric oscillator we find integrals
of motion which satisfy the commutation relations of q--oscillator
\cite{bie}, \cite{mc}, i.e. the quantum Heisenberg--Weyl group
is shown to be the dynamical group of this oscillator. In \cite{sol1},
\cite{sol2} the possibilities to interprete the q--oscillator as nonlinear
oscillator with specific exponential dependence of its frequency  on the
amplitude has been suggested. Using the construction of even and odd
coherent states \cite{dod74} interpreted in quantum optics as
Schr\"odinger cat states \cite{sto} we discuss the properties of these
states for the Husimi parametric oscillator.

For symplicity of presentation we take $~f(t)=0$. Then the time--dependent
operator
\begin{equation}
A=\frac {i}{\sqrt 2}[\varepsilon (t)p-\dot \varepsilon (t)x],
\end{equation}
where
\begin{equation}
\ddot \varepsilon (t)+\omega ^{2}(t)\varepsilon (t)=0,~~~~~~
\varepsilon (0)=1,~~~~~~\dot \varepsilon (0)=i
\end{equation}
is the integral of motion \cite{mal70} satisfying the commutation relation
\begin{equation}
[A,~A\dag ]=1.
\end{equation}
One can easily check that the time--dependent operator
\begin{equation}
A_{q}=A\sqrt {\frac {\sinh \lambda A\dag A}{A\dag A\sinh \lambda }},
\end{equation}
where $~\lambda =\ln q$, is the integral of motion of the parameteric
oscillator satisfying the commutation relation
\begin{equation}
A_{q}A_{q}\dag -qA_{q}\dag A_{q}=q^{-A\dag A}.
\end{equation}
It means that operators $~A_{q}~(A_{q}\dag )$ corresponding to q--oscillators
due to these commutation relations determine the dynamical symmetry of the
parametric oscillator and all the states of the oscillator are the basis of
the irreducible representation of q--deformed Heisenberg--Weyl group.

\section{Correlated and Squeezed State}
It was shown in \cite{mal70} that some Husimi packet solutions may be
interpreted as coherent states \cite{gla63} since they are eigenstates of
the operator $~A$ (2) of the form
\begin{equation}
\Psi _{\alpha }(x,t)=\Psi _{0}(x,t)\exp \{-\frac {|\alpha |^{2}}{2}-
\frac {\alpha ^{2}\varepsilon ^{*}(t)}{2\varepsilon (t)}
+\frac {{\sqrt 2}\alpha x}{\varepsilon}\},
\end{equation}
where
\begin{equation}
\Psi _{0}(x,t)=\pi ^{-1/4}\varepsilon (t)^{-1/2}
\exp \frac {i\dot \varepsilon (t)x^{2}}{2\varepsilon (t)}
\end{equation}
is analog of the ground state of the oscillator and $~\alpha $ is a
complex number. The variances of the position and momentum of the
parametric oscillator in the state (8) are
\begin{equation}
\sigma _{x}=\frac {|\varepsilon (t)|^{2}}{2},~~~~~~\sigma _{p}
=\frac {|\dot \varepsilon (t)|^{2}}{2},
\end{equation}
and the correlation coefficient of the position and momentum has
the value corresponding to minimization of the Schr\"odinger uncertainty
relation \cite{schrod30}
\begin{equation}
\sigma _{x}\sigma_{p}=\frac {1}{4}\frac {1}{1-r^{2}}.
\end{equation}
Due to this the state (7) is a correlated coherent state introduced in
\cite{kur}. For different dependences of frequency on time there may
appear phenomenon of squeezing, i. e.
\begin{equation}
\sigma _{x}<1/2.
\end{equation}
The orthonormal number states of parametric oscillator
\begin{equation}
\Psi _{n}(x,t)
=\Psi _{0}(x,t)\frac {1}{\sqrt {n!}}(\frac {\varepsilon ^{*}}
{2\varepsilon })^{n/2}H_{n}(\frac {x}{|\varepsilon (t)|})
\end{equation}
are eigenstates og the integral of motion $~A\dag A$. For the Husimi
oscillator the q--coherent states \cite{bie} may be easily constructed
as normalized eigenstates of the integral of the motion $~A_{q}$ (5)
in the form
\begin{equation}
\Psi _{\alpha }^{(q)}=N_{q}\sum _{n=0}^{\infty }\frac {\alpha ^{n}}
{\sqrt {[n!]}}\Psi _{n}(x,t),
\end{equation}
where $~[n!]=~[n][n-1]\cdots [1],~[0!]=1$ and $~[n]=
{}~\sinh \lambda n/\sinh \lambda .$ The normalization constant is
\begin{equation}
N_{q}=(\sum _{n=0}^{\infty }\frac {|\alpha |^{2n}}{[n!]})^{-1/2}
\end{equation}
The function (13) satisfies the Schr\"odinger equation (1), and it is
eigenfunction of the q--oscillator operator $~A_{q}$. In \cite{sol1},
\cite{sol2} it was shown that if the q--oscillator describes a
nonlinearity of classical electromagnetic field vibrations
(q--nonlinearity) at high amplitudes the blue shift effect must exist.
It is the effect of shifting the light frequency versus the light
intensity which may serve for evaluating an upper limit for the
parameter of nonlinearity $~\lambda .$

\section{Squeezed Schr\"odinger Cats}
Another normalized solution to the Schr\"odinger equation (1)
\begin{equation}
\Psi _{\alpha m}(x,t)=2N_{m}\Psi _{0}(x,t)\exp \{-\frac {|\alpha |^{2}}
{2}-\frac {\varepsilon ^{*}(t)\alpha ^{2}}{2\varepsilon (t)}\}\cosh
\frac {{\sqrt 2}\alpha x}{\varepsilon (t)},
\end{equation}
where
\begin{equation}
N_{m}=\frac {\exp (|\alpha |^{2}/2)}{2\sqrt {\cosh |\alpha |^{2}}}
\end{equation}
is the even coherent state \cite{dod74} (the Schr\"odinger cat male state).
The odd coherent state of the parametric oscillator (Schr\"odinger cat
female state)
\begin{equation}
\Psi _{\alpha f}(x,t)=2N_{f}\Psi _{0}(x,t)\exp \{-\frac {|\alpha |^{2}}
{2}-\frac {\varepsilon ^{*}(t)\alpha ^{2}}{2\varepsilon (t)}\}\sinh
\frac {\sqrt {2}\alpha x}{\varepsilon (t)},
\end{equation}
where
\begin{equation}
N_{f}=\frac {\exp (|\alpha |^{2}/2)}{2\sqrt {\sinh |\alpha |^{2}}}
\end{equation}
satisfies the Schr\"odinger equation (1) and is the eigenstate of the
integral of motion $~A^{2}$ (as well as the even coherent state) with
the eigenvalue $~\alpha ^{2}.$ The even and odd coherent states are
ortogonal each to other. A multimode generalization of the even and odd
coherent states has been introduced in \cite{ans94}. The properties of
multimode Schr\"odinger cat states, the coherent components of which are
subjected to parametric excitation, are discussed in \cite{nikon}. The
even and odd coherent states are the partial case of a set of nonclassical
states discussed in \cite{nieto}. They may be used as alternative to
squeezed vacuum states in gravitational wave antenna to improve its
sensitivity \cite{marga}.

The multimode generalization of the Husimi parametric oscillator has been
considered in \cite{tri73}. The linear in position and momentum integrals
of motion have been constducted and generalized correlated states
\cite{sudar} in the form of Gaussian packets have been found explicitly.
It is interesting to note that for multidimensional parametric oscillator
the linear integrals of motion of the type $~A$ (2) and their quadratic
forms constitute the Lie algebra of the inhomogeneous symplectic group.
Thus, the solutions to Schr\"odinger equation for the multimode parametric
oscillator form the basis for irreducible representation of this symplectic
group. The remark made for one--mode case on the existence of q--deformed
dynamical algebra may be extended to the multimode parametric oscillator,
too. Thus, the solutions of the Schr\"odinger equation for multimode
parametric oscillator may be also considered as a basis of the irreducible
representation for a q--deformed symplectic group.

The construction of the even and odd coherent states made in \cite{dod74}
used the representation of the inversion group for which even states realize
the symmetric irreducible representation and odd states realize the
antisymmetric one. Generalization of this construction to higher
crystallographic groups for introducing the crystallized Schr\"odinger cat
states is done in \cite{costan94}.

\section{Mulivariable Hermite Polynomials}
For Husimi parametric forced oscillator the transition amplitude between
its energy levels has been calculated as overlap integral of two generic
Hermite polynomials with a Gaussian function (Frank--Condon factor) and
expressed in terms of Hermite polynomials of two variables \cite{mal70}.
For N--mode parametric oscillator the analogous amplitude has been
expressed in terms of Hermite polynomials of 2N variables, i. e. the
overlap integral of two generic Hermite polynomials of N variables with
a Gaussian function (Frank--Condon factor for a polyatomic molecule) has
been evaluated in \cite{tri73}. The corresponding result uses the formula
$~({\bf n}=~n_{1},~n_{2},\cdots ~n_{N},~~~~~{\bf m}=~m_{1},~m_{2},\cdots
{}~m_{N},~~~~~m_{i},~n_{i}=~0,~1,\cdots)$
\begin{equation}
\int H_{{\bf n}}^{\{R\}}H_{{\bf m}}^{\{r\}}(\Lambda {\bf x}+{\bf d})
\exp (-{\bf x}m{\bf x}+{\bf c}{\bf x})d{\bf x}=\frac {\pi ^{N/2}}
{\sqrt {\det m}}\exp (\frac {1}{4}{\bf c}m^{-1}{\bf c})
H_{{\bf mn}}^{\{\rho \}}({\bf y}),
\end{equation}
where the symmetric 2N$\times $2N--matrix
\begin{equation}
\rho=\left( \begin{array}{clcr}R_{1}&R_{12}\\
\widetilde R_{12}&R_{2}\end{array}\right)
\end{equation}
with N$\times $N--blocks $~R_{1},~R_{2},~R_{12}$ is expressed in terms of
N$\times $N--matrices $~R,~r,~m$ and a N$\times $N--matrix $~\Lambda $
in the form
\begin{eqnarray}
R_{1}&=&R-\frac {1}{2}Rm^{-1}R\nonumber\\
R_{2}&=&r-\frac {1}{2}r\Lambda m^{-1}\tilde \Lambda r\nonumber\\
\widetilde R_{12}&=&-\frac {1}{2}r\Lambda m^{-1}R.
\end{eqnarray}
Here the matrix $~\widetilde \Lambda $ is transposed matrix $~\Lambda $
and $~\widetilde R_{12}$ is transposed matrix $~R_{12}.$ The 2N--vector
$~{\bf y}$ is expressed in terms of N--vectors $~{\bf c}$ and $~{\bf d}$
in the form
\begin{equation}
{\bf y}=\rho ^{-1}\left( \begin{array}{c}{\bf y}_{1}\\
{\bf y}_{2}\end{array}\right ),
\end{equation}
where the N--vectors $~{\bf y}_{1}$ and $~{\bf y}_{2}$ are
\begin{eqnarray}
{\bf y}_{1}&=&\frac {1}{4}(Rm^{-1}+m^{-1}R){\bf c}\nonumber\\
{\bf y}_{2}&=&\frac {1}{4}(r\Lambda m^{-1}
+m^{-1}\tilde \Lambda r){\bf c}+{\bf d}.
\end{eqnarray}
The multivariable Hermite polynomials describe the photon distribution
function for the multimode mixed and pure correlated light \cite{olga},
\cite{md94}, \cite{dodon94}. The nonclassical state of the light may be
created due to nonstationary Casimir effect \cite{jslr}, and the
multimode Husimi oscillator is the model to describe the behaviour of
the squeezed and correlated photons.

\end{document}